# Studies of PHP with CASCO code and its experimental validation

Vadim S. Nikolayev[1*], Slaven Bajić[2], Guillaume Boudier[2], Eric Blondé[3], Typhaine Coquard[3], Mauro Abela[1,4], Mauro Mameli[4], Sauro Fillipeschi[4]

[1]*SPEC CEA Paris-Saclay, and Université Paris-Saclay, Gif-sur-Yvette, France*
[2]*Service de thermique, Centre National d'Études Spatiales (CNES), Toulouse, France*
[3]*Airbus Defence and Space, Toulouse, France*
[4]*Department of Energy, Systems, Land and Construction Engineering, University of Pisa, Italy*
**Corresponding author email address: vadim.nikolayev@cea.fr*

## Abstract

We discuss here two major issues related to the steady functioning of the pulsating (oscillating) heat pipe (PHP): the effect of the surface properties and stopovers. They are studied with the CASCO simulation software (Code Avancé de Simulation du Caloduc Oscillant: Advanced PHP Simulation Code in French) version 4. Its experimental validation against two different prototypes is presented. The first is used also to study the effect of the nucleation barrier (the wall superheating necessary for the bubble nucleation) that reflects the wall wettability and roughness. An optimal value of the nucleation barrier is found where the thermal resistance achieves a minimum for a given evaporator power. The functioning regime is continuous showing pressure waves propagating along all the PHP channel. The stopover regime is observed both for small and large barriers. The second experimental setup (PHP Smart Loop) is used to study the stopover regime. It is found that it is characterized by a chaotically repeating sequence of fast pressure growth (corresponding to oscillations) followed by a slower pressure decay during a stopover. The decrease of the thermal resistance with heating load is explained by a decrease of the stopover time caused by a faster liquid film shrinking.

*Keywords*: Pulsating heat pipe; Simulation with CASCO; Regime map; Nucleation; Liquid films

## 1. Introduction

The pulsating (or oscillating) heat pipe (PHP) is a simple capillary tube bent into branches meandering between heat source and sink and partially filled with a two-phase working fluid. A pattern of multiple vapor bubbles separated by liquid plugs forms inside the tube. A combination of the latent heat transfer produced by the phase change and the convective boiling and condensation leads to a high performance of the PHP. Combined with its simplicity (thus high reliability and low cost), it makes PHP to be of high industrial potential. Last decades, researchers have extensively studied PHP [1] with dozens of research articles published each year in leading scientific journals. However, the PHP functioning is intrinsically non-stationary and even chaotic. Simple tools for designing the PHP are still absent and the only way to predict its functioning is the numerical simulation [2] .

In this article, the PHP parametrical study is performed for the steady operation state with the CASCO simulation software (Code Avancé de Simulation du Caloduc Oscillant: Advanced PHP Simulation Code in French); its experimental validation is also presented.

One of the most critical issues of the PHP functioning is the strong dependence on the state of internal surface of PHP channels. There were many attempts to improve the PHP performance by artificially varying the surface state. In many studies [3] –[5] , the roughness was changed, while in some others [6] –[8] , the wetting properties were modified. In some cases, the properties were modified only in some sections (e.g. in evaporator [7] ). At the end, the micrometric roughness modification leads to variation of the static (contact angle) as well as dynamic (pinning of the contact lines) wetting properties [9] . Therefore, we will reason in terms of wettability variation that can be summarized in Table 1. The data are sometimes contradictory, but most researchers agree that the

**Table 1**. Effect of wettability on PHP.

| | startup | $R_{th}$ at low $P_e$ | $R_{th}$ at high $P_e$ | $P_{dryout}$ |
|---|---|---|---|---|
| wettable | earlier | higher | lower | higher |
| non-wettable | later | lower | higher | lower |

A wettable internal surface is generally better, i.e., provides earlier oscillation startup, higher dryout threshold $P_{dryout}$ and smaller thermal resistance $R_{th}$ (probably except at small evaporator heat powers $P_e$, which can be a consequence of stronger contact line pinning at higher wettability [10] ). In CASCO, there is no direct way to account for the surface state as it is a 1D code. However, it is well established that the surface state is directly related to the wall superheating required for the vapor to the nucleation barrier $\Delta T_{nucl}$, which represents the bubble incipience. This quantity is called also the superheating of the onset of nucleate boiling. For the non-wettable surfaces $\Delta T_{nucl}$ is smaller than for wettable surfaces [11] .

Another issue in the field of PHP are the stopovers and restarting of PHP after them. This issue is critical as it leads to the criterion of the PHP startup and thus operation limitation. In this paper we discuss this issue too by studying the stopovers both experimentally and with CASCO simulations.

## 2. Copper PHP and its simulation

The PHP of $N_{turn}$ =16 turns (Figure 1) is made of the copper tube ($r = 0.5$ mm internal and $r_{ext} = 1.5$ mm external radius) filled with the Novec® 7100 (HFE7100) fluid (FR=0.55). On each side of the PHP there are two similar spreaders. Each of them consists of a pair of blocks made of aluminum alloy 6061T6 with half-cylindrical notches that enclose the copper tubes. The thermal paste is applied between the spreader blocks and the copper tubes to improve the thermal contact. The block near the feedback loop serves as the condenser. Its temperature is regulated by a cold plate with internally pumped coolant flow, attached to the lower side of the condenser block. The cold plate temperature is 20°C. The other block acts as the evaporator spreader, to which heating power is applied using Kapton® polyimide heaters.

### 2.1. Simulation details

The practical implementation of numerical simulation requires projection of the PHP meandering tube to a straight 1D axis (abscissa $x$) with periodical boundary conditions at its ends [2] . The lateral boundary conditions are different for evaporator (heat input), adiabatic (heat exchange with the environment only) and condenser sections (constant temperature at the internal tube walls). At each $x$, the fluid can be in one of three states: liquid plug, vapor bubble with a liquid film covering the wall surface, or without it (also referred to as the dry spot). Fluid properties are automatically extracted from the RefProp database [12] .

The temporal evolution of the internal wall temperature $T_w$ obeys the equation

$$\rho_w c_w \frac{\partial T_w}{\partial t} = \lambda_w \frac{\partial^2 T_w}{\partial x^2} + \frac{2\pi}{S_w}\left\{ \begin{matrix} r_{ext}\, q_{se}(x) - r\, q_f(x) & \text{if } x \in \text{evaporator} \\ -r_{ext} q_{air}(x) - r\, q_f(x) & \text{if } x \in \text{adiabatic sect.} \end{matrix} \right\} \quad (1)$$

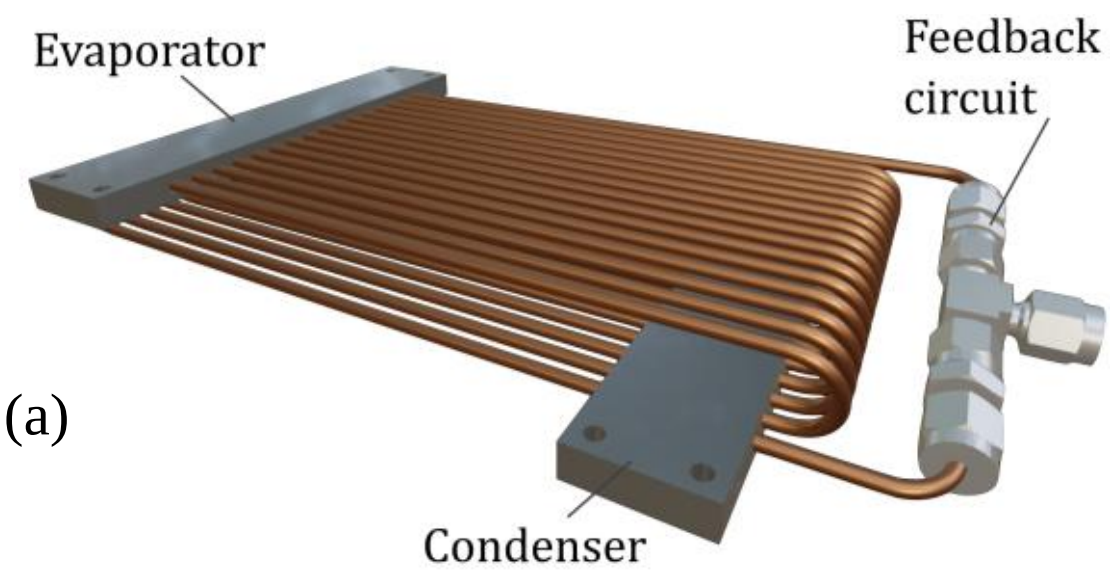

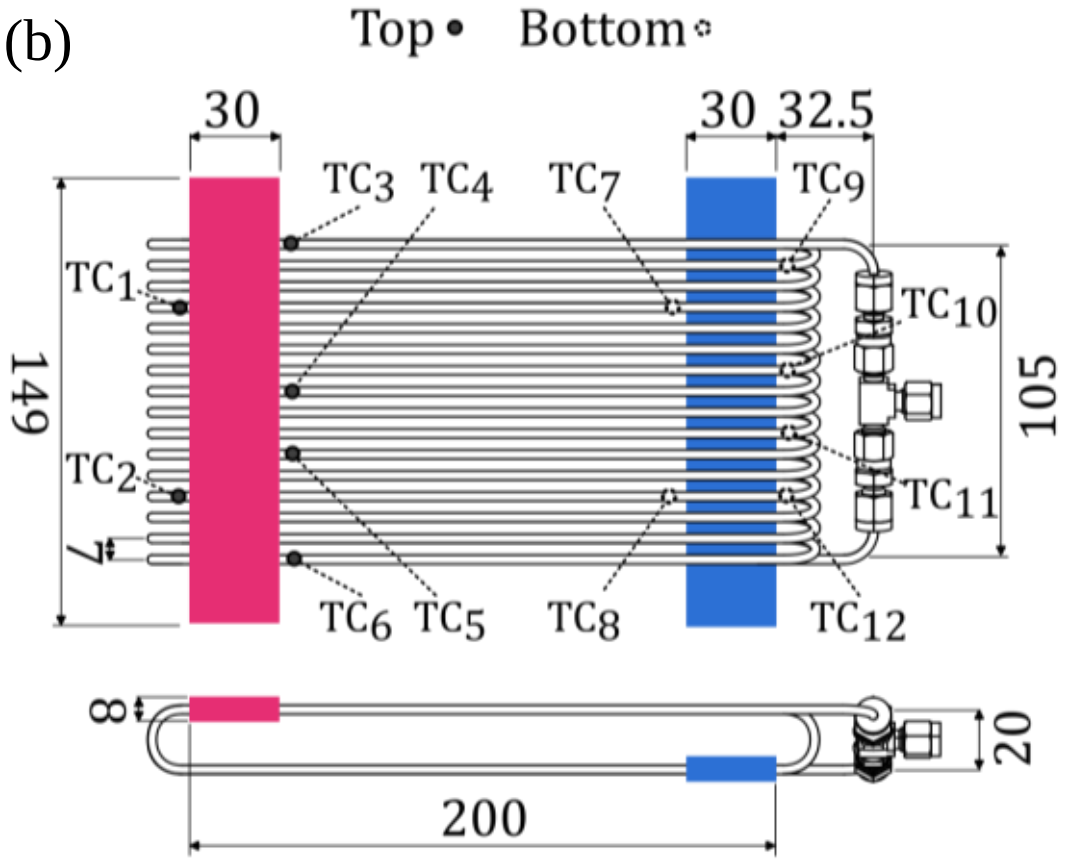

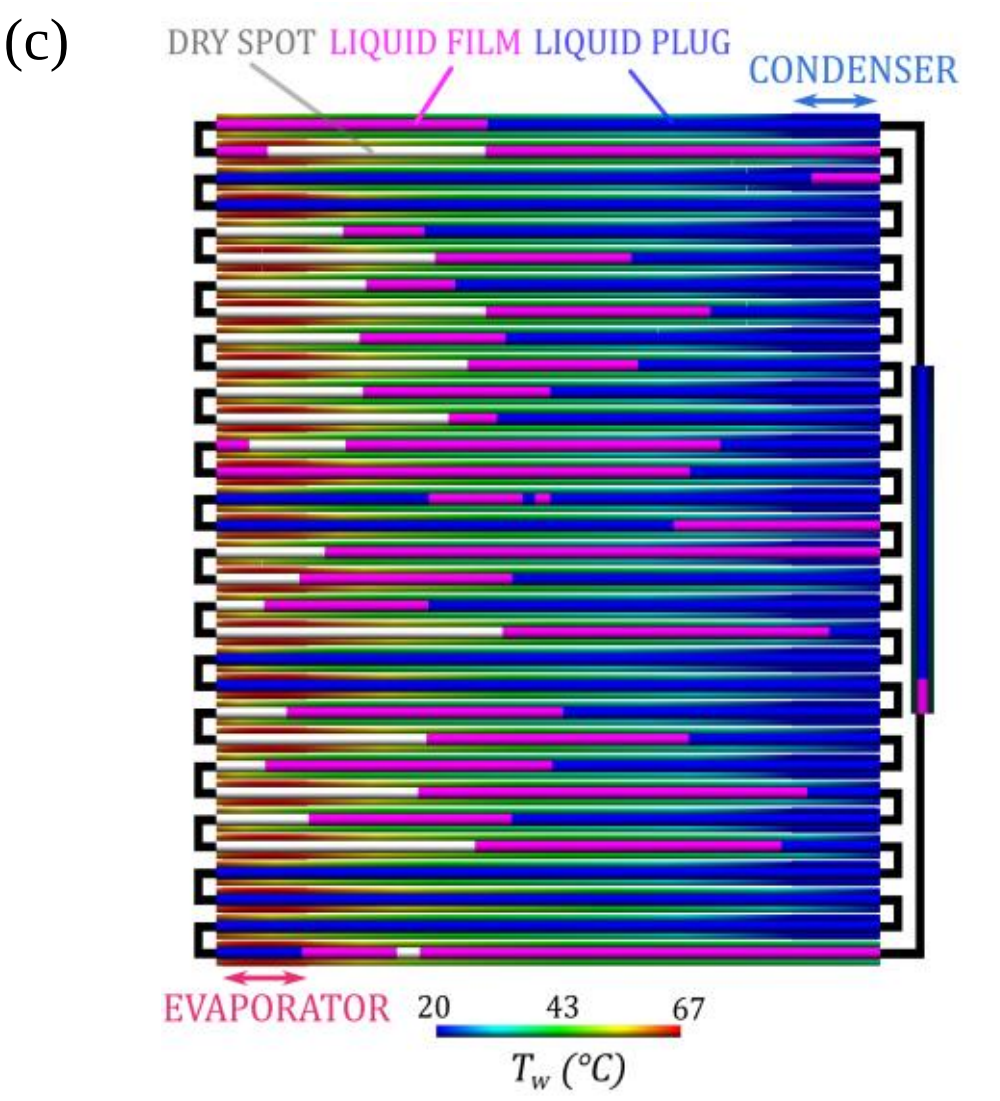

**Figure 1.** Copper PHP geometry (a), sizes (b) and (c) an example of visualization of the CASCO fluid phase distributions and wall temperature at $P_e$= 70 W, $T_c$ = 19.5 °C and $t$ = 1200 s. The evaporator end condenser sections are located in each second branch by alternating between them. The temperature scale shows the spatial distribution of the tube temperature $T_w$.

where $S_w = \pi\,(r_{ext}^2 - r^2)$ is the tube cross-section; $\rho_w$, $c_w$ and $\lambda_w$ are the wall material density, specific heat and thermal conductivity, respectively. The wall-fluid heat flux $q_f$ is determined with other CASCO equations [13] , [14] . Additionally, CASCO accounts for the heat flux

$$q_{air}(x) = h_{air}\,[T_w(x) - T_{air}] \tag{2}$$

of the losses to environment from the PHP tubes in adiabatic sections. Here, $T_{air}$ is the temperature of environment and $h_{air}$ is the respective heat exchange coefficient. In the evaporator sections, the tubes receive the heat flux

$$q_{se}\,(x) = h_{se}[T_s - T_w(x)] \tag{3}$$

from the evaporator spreaders at the temperature $T_s$ ; $h_{se}$ is the thermal conductance of corresponding contact. Eq. (1) is solved in the evaporator and adiabatic sections; in the condenser, $T_w = T_c$ is imposed.

The evaporator spreader block is highly conductive and thus isothermal; its thermal mass is denoted $C_s$. Its temperature obeys the energy balance

$$C_s \frac{dT_s}{dt} = P_e - P_{air} - 2\,\pi\, r_{ext} \int_0^{L_t} \begin{Bmatrix} q_{se}(x), & \text{if } x \in \text{evap.} \\ 0, & \text{otherwise} \end{Bmatrix} dx, \tag{4}$$

where $L_t = N_{turn}\,(L_e + L_{a1} + L_c + L_{a2}) + L_{fb}$ is the total PHP tube length; the section lengths are shown in Table 2. The spreader is heated due to the evaporator power $P_e$. The heat power $P_{air}$ corresponds to the heat losses from the spreader to the environment,

$$P_{air} = H_{air}\,(T_s - T_{air}), \tag{5}$$

where $H_{air}$ is the integral heat loss coefficient.

The liquid film thickness $\delta$ strongly influences the overall heat transfer and fluid dynamics [15] and should be specified as a constant initial parameter in the input CASCO file together with other parameters of Table 2. However, from the physical point of view, it depends on the velocity of receding meniscus that deposits the film according to the formula [16]

$$\delta = \frac{1.34\, r Ca^{2/3}}{1+3.35\, Ca^{2/3}}, \tag{6}$$

where $Ca$ is the capillary number corresponding to the plug speed. To obtain the average $\delta$ value, we use the following iterative procedure. Initially, a seed $\delta$ value is chosen. One performs an iteration step that begins with running CASCO. Next, one determines the average plug speed $\bar{V}_{rms}$. To obtain it, one first computes the root-mean-square of all instantaneous plug velocities denoted $V_{rms}$. Then they are arithmetically averaged over the time (which is signaled by overbar) in the steady functioning regime (in practice, over last 50 s out of simulated 1200 s); the capillary number writes $Ca = \mu\, \bar{V}_{rms}/\sigma$. The temperature averaged over all the evaporator sections $T_{ref}$ is used to calculate the shear viscosity $\mu$ and the surface tension $\sigma$ with RefProp. The same $T_{ref}$ is used for calculation of the fluid parameters for the next iteration step, where the $\delta$ value determined with (6) is used. The iterations are repeated until the input and output $\delta$ value coincide within the required accuracy (here, 5%).

**Table 2.** Input CASCO parameters employed for the copper PHP simulation.

| Parameter and its notation | Value |
|---|---|
| Evaporator section length $L_e$ | 30 mm |
| Condenser section length $L_c$ | 30 mm |
| Adiabatic section lengths $L_{a1}$, $L_{a2}$ | 204 mm, 202 mm |
| Feedback section length $L_{fb}$ | 116 mm |
| Tube bend radius | 10 mm |
| Time step | 0.2 ms |
| Wall element length | 1 mm |
| Liquid element length | 1 mm |
| Nucleated bubble length $L_{nucl}$ | 0.1 mm |
| Minimum distance of nucleated bubble from the plug end $L_{nucl,min}$ | 5 mm |
| Bubble deletion threshold $L_v^{thr}$ | 0.01 mm |
| Liquid plug deletion threshold $L_l^{thr}$ | 2 mm |
| Tube wall heat capacity $c_w$ | 389.3 J.kg$^{-1}$.K$^{-1}$ |
| Tube wall density $\rho_w$ | 8915 kg.m$^{-3}$ |
| Tube wall thermal conductivity $\lambda_w$ | 398.4 W.m$^{-1}$.K$^{-1}$ |
| Evap. spreader thermal mass $C_s$ | 250 J.K$^{-1}$ |
| Thermal conductance between evaporator spreader and tubes $h_{se}$ | 3800 W.m$^{-2}$.K$^{-1}$ |
| Tube heat loss coefficient $h_{air}$ | 0.2 W.m$^{-2}$.K$^{-1}$ |
| Spreader heat loss coefficient $H_{air}$ | 0.05 W.K$^{-1}$ |
| Temperature of environment $T_{air}$ | 19.5°C |

Since the PHP behavior is sensitive to $\delta$, its variation was used to evaluate the uncertainty of the thermal resistance, i. e. the simulation error bar. By altering $\delta$ by up to 15%, the deviation in PHP thermal resistance did not exceed 5% in all the investigated cases.

### 2.2. CASCO validation against the measurements of copper PHP

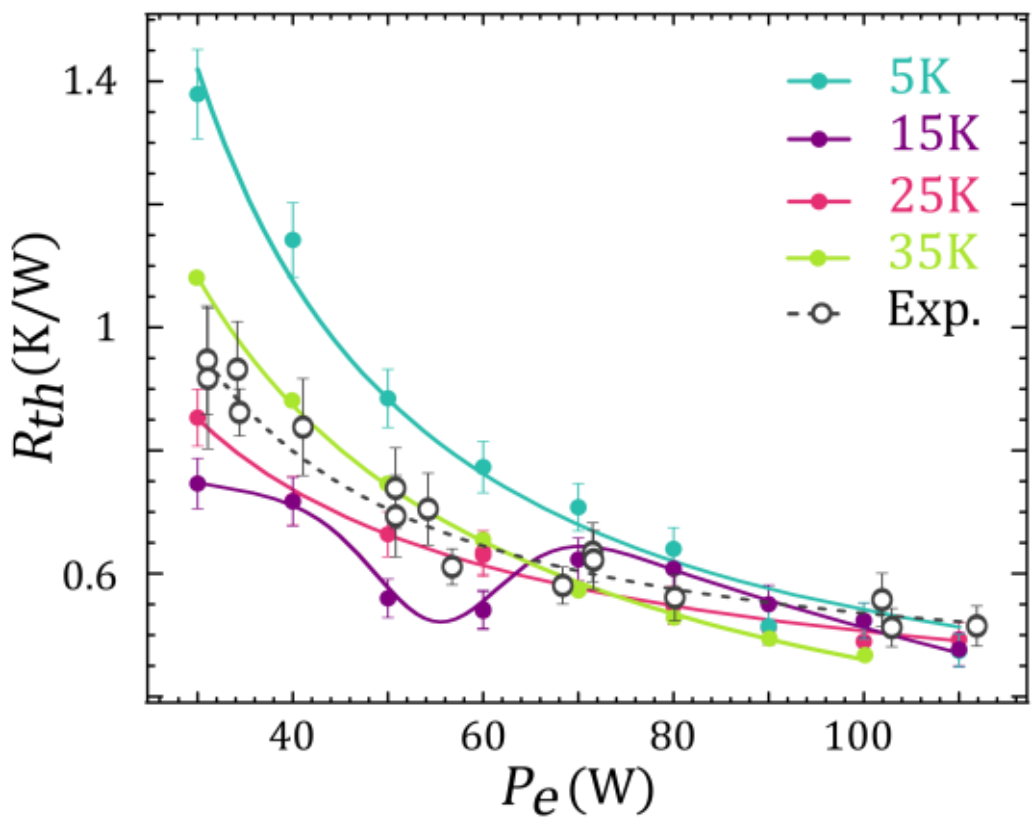

**Figure 2.** Experimentally measured thermal resistance for the copper PHP and CASCO simulations for different $\Delta T_{nucl}$.

We validate CASCO by using the PHP thermal resistance defined as

$$R_{th} = \frac{\bar{T}_e - T_c}{P_e}. \quad (7)$$

The temperature was measured with the thermocouples $TC_1$-$TC_6$ located on the tubes next to the evaporator spreader and $TC_7$-$TC_{12}$ near that of condenser (Figure 1b). Thermocouples were also used to measure the spreader temperature that is assumed to be equal to that of the internal tube walls $T_c$. To provide more accurate comparison by avoiding the effect of the spreader thermal connection to the tubes, the average evaporator temperature is defined as

$$\bar{T}_e = \frac{1}{6}\sum_{i=1}^{6} T_i, \quad (8)$$

where $T_i$ is the temperature measured by the thermocouple $TC_i$. The uncertainty associated with $R_{th}$ is estimated by propagating the standard deviation from the mean temperature: the main source of uncertainty is the temperature difference between measurements at different positions.

To prepare the main CASCO runs, an auxiliary simulation was performed with CASCO for the empty PHP case to tune the model parameters of the solid structure $h_{air}$, $H_{air}$ and $h_{se}$. Their values found from the best fit of the experimental data with the CASCO simulations for the empty PHP are shown in Table 2.

The simulations replicate the conditions from the experimental campaign ($P_e$ ranging from 30 to 110 W). Since the condenser temperature in the experimental investigations varies with heating power due to a thermal resistance between the cooling plate and the condenser spreader, $T_c$ variation with $P_e$ in the simulations is taken from the experimental measurements.

Figure 2 shows an $R_{th}$ agreement obtained for $\Delta T_{nucl}$=25 K. It is difficult to measure this value experimentally inside a tube. Similar $\Delta T_{nucl}$ values were measured [17] ,[18] for smooth and wettable surface with other fluorocarbon fluids; we expect them to be similar to HFE7100 used here. Note that the absolute value of the CASCO $\Delta T_{nucl}$ parameter can be different from the actual physical value. The nucleated bubble in CASCO is a thin disk (of the thickness $L_{nucl}$ shown in Table 2) rather than a small sphere.

### 2.3. Study of PHP oscillation regimes

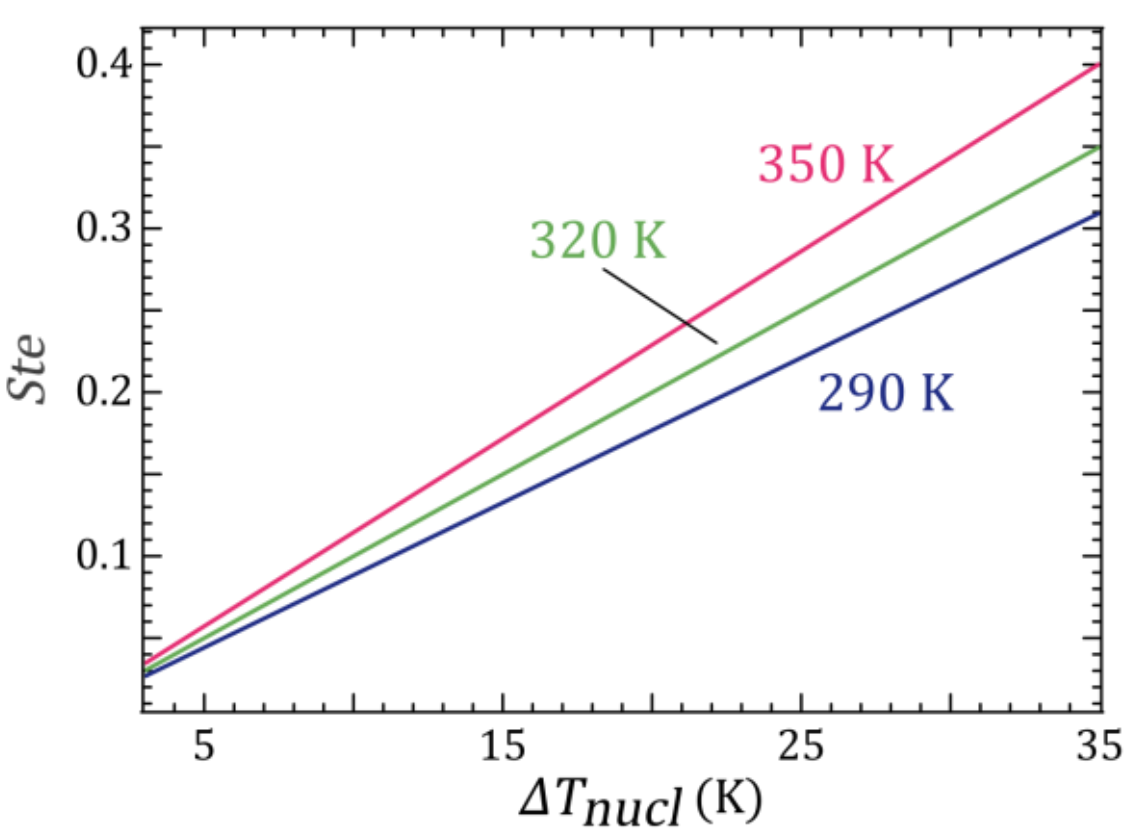

**Figure 3.** Stefan number as a function of the nucleation barrier for different $T_{ref}$.

The dependence on $\Delta T_{nucl}$ is studied in terms of the dimensionless Stefan number $Ste = \frac{c_l}{\mathcal{L}} \Delta T_{nucl}$ containing the latent heat $\mathcal{L}$ and the liquid heat capacity $c_l$. The choice of $Ste$ to make $\Delta T_{nucl}$ dimensionless is justified by a strong link of bubble nucleation to the liquid evaporation. Both $c_l$ and $\mathcal{L}$ are evaluated for each steady PHP state and are based on $T_{ref}$ (Figure 3) that is determined as explained in section 2.1.

The PHP functioning as a function of $Ste$ is analyzed in Figure 4. The dots correspond to the simulation data and the lines are their fits. The variation of $R_{th}$ (Figure 4a) is non-monotonous at each fixed $P_e$, which suggests a change in the functioning mode. The functioning is more efficient in a middle range of $Ste$, especially well defined at small $P_e$ values. The $\bar{V}_{rms}$ variation with $Ste$ (Figure 4b) shows a similar tendency: the oscillations are faster on average in the middle range, which should thus correspond to a particular regime. Its nature can be found by plotting the frequency of occurrence of a given $V_{rms}$ value in Figure 5. One can see that all the distributions corresponding to the middle range (shown in yellow) have a cut-off value, never lower than 0.4 m/s, which means that the oscillations are continuous. Outside the middle $Ste$ range, one finds a non-negligible occurrence of small (and even zero) velocities, which means the PHP stopovers. One can thus plot the regime map (Figure 4c) depicting the continuous and stopover oscillation regimes by using 0.4 m/s as a threshold speed.

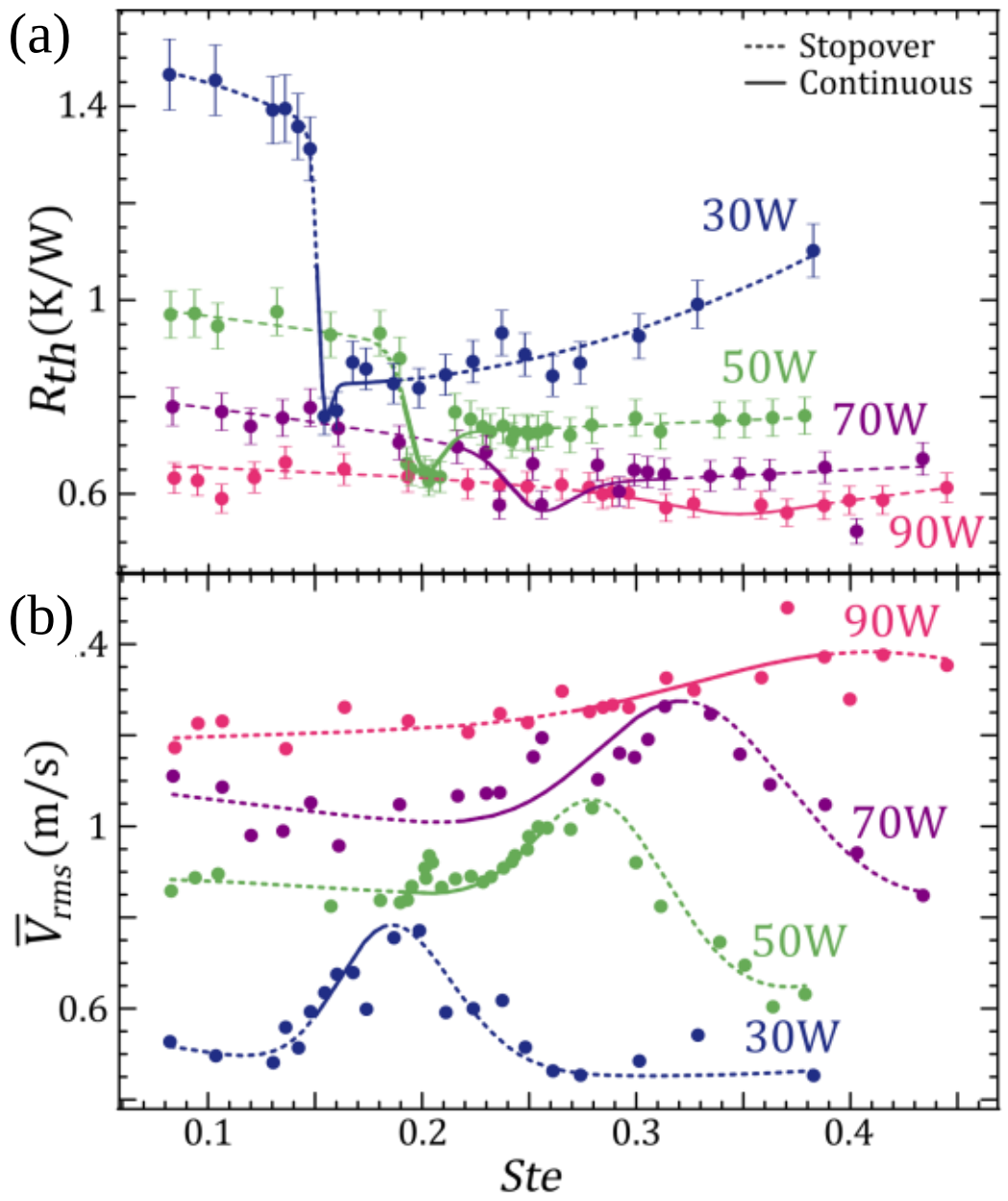

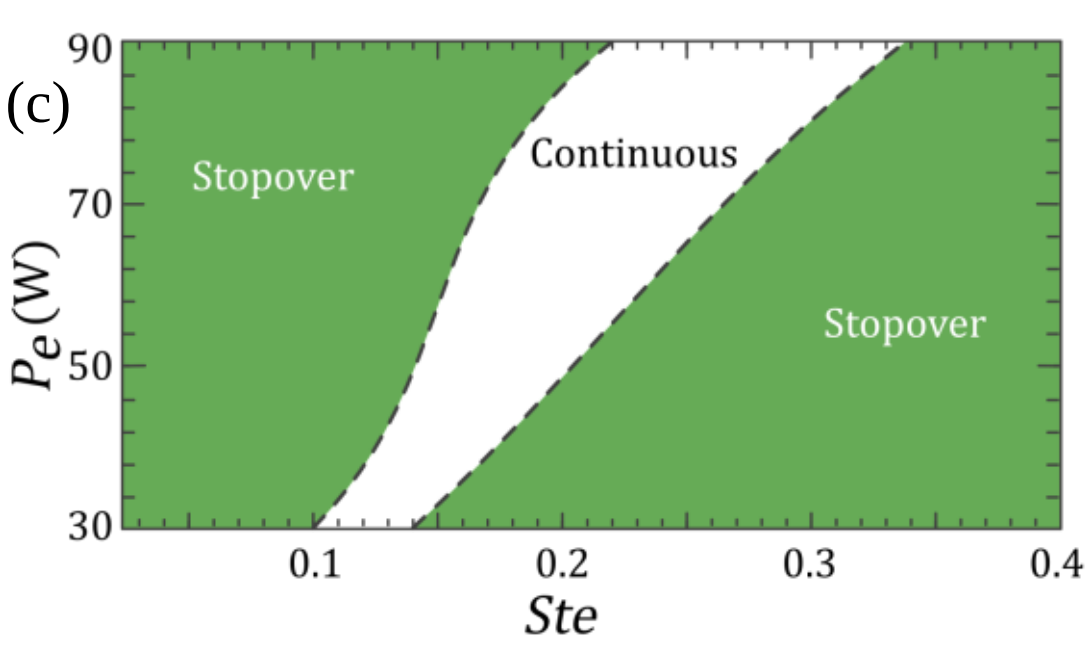

**Figure 4.** Simulated dependencies on the nucleation barrier expressed with $Ste$ for different heating powers $P_e$: (a) thermal resistance, (b) plug speed and (c): map of regimes of oscillation.

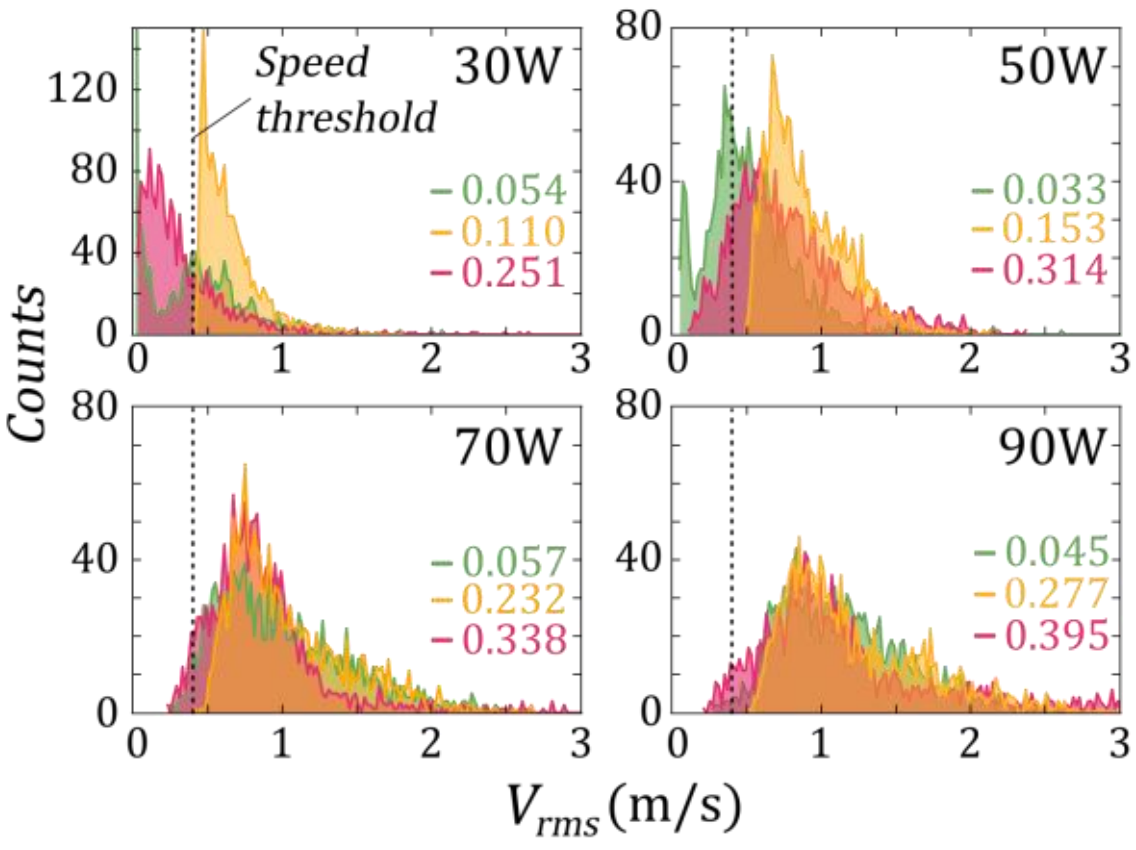

**Figure 5.** $V_{rms}$ occurrence distributions at four $P_e$ and different $Ste$ values. The vertical dashed line shows the speed threshold of 0.4 m/s.

Both regimes are represented in Figures 4a,b for each $P_e$ value as solid and dashed portions of the corresponding line.

## 3. Study of stopover regime

This study is performed by using the "Smart Loop" apparatus (Figure 6), which consists of four 765 mm glass tubes connected at the corners by four brass joints denoted A–D that contain the temperature and pressure sensors. The tubes are made of silica glass ($r$ =1 mm, $r_{ext}$ =2.5 mm). The electrically conductive transparent 4 cm-long ring patches separated by 4 mm spaces are made of ITO deposited on the external tube surface. An evaporator electrically-heated section of arbitrary length can be created by connecting neighboring patches. The thermal bath-cooled condenser sections are made with aluminum spreaders that sandwich the glass tubes. The Smart Loop is thus equivalent to a horizontal PHP with a number of turns equal to the number of evaporator sections, which can be modulated thus justifying the Smart Loop name. In the considered case it was filled with ethanol (FR=0.5) and $N_{turn}$ = 11. CASCO was used to simulate [19] the Smart Loop functioning; the CASCO input parameters are shown in Table 3. Only the parameters different from those of Table 2 are mentioned. The management of heat losses is slightly different with respect to those described in section 2. First, the spreader is absent in Smart Loop and the heat losses $q_{air}(x)$ to the environment act not only in adiabatic but also in the evaporator section. Second, the temperatures of inner and outer tube walls cannot be assumed equal in this case because of the low glass conductivity; a model of the radial heat conduction [19] is introduced.

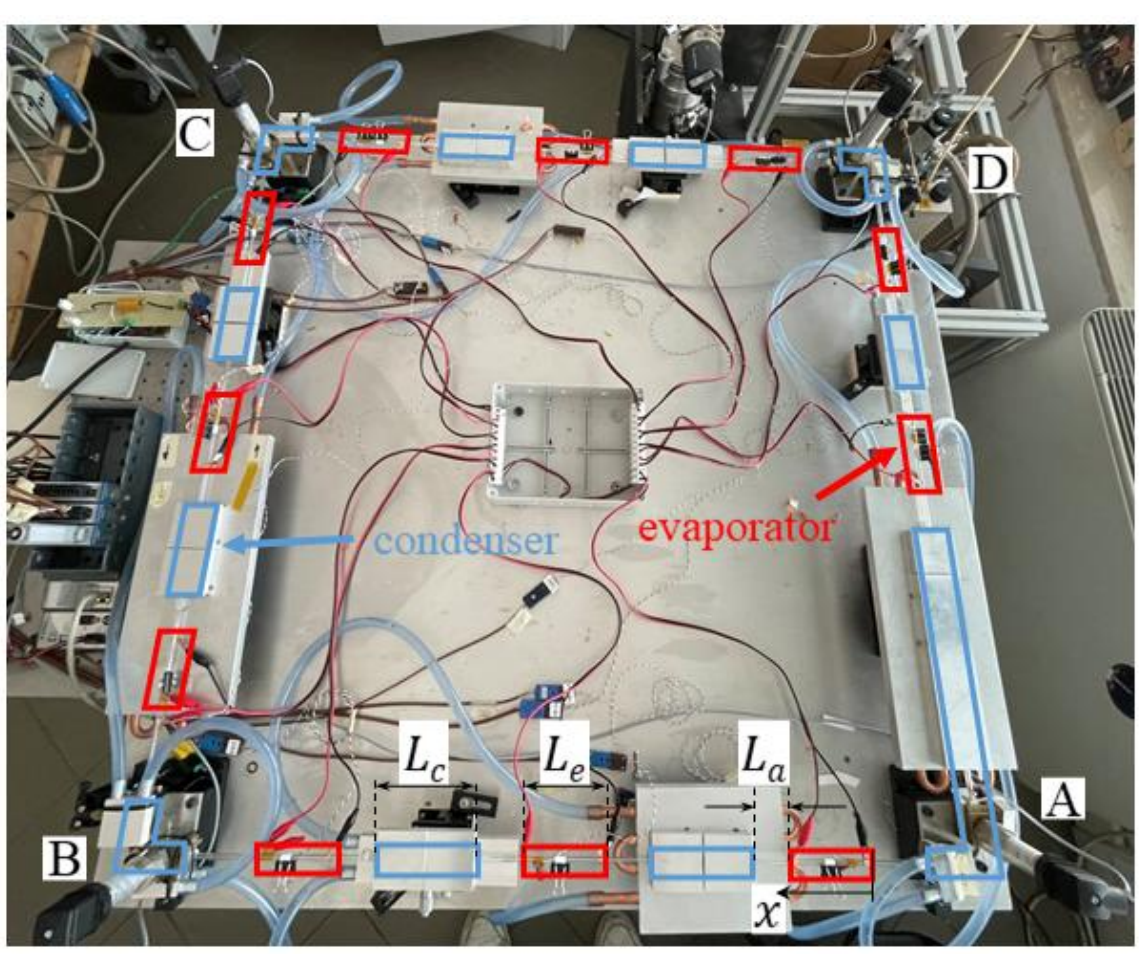

**Figure 6.** Picture of the "Smart Loop" with the indicated evaporator (red) and condenser (blue) sections [19] .

**Table 3.** Smart Loop simulation parameters.

| Parameter and its notation | Value |
|---|---|
| Evaporator section length $L_e$ | 84 mm |
| Condenser section length $L_c$ | 100 mm |
| Adiabatic section lengths $L_{a1}$, $L_{a2}$ | 50 mm, 50 mm |
| Feedback section length $L_{fb}$ | 236 mm |
| Time step | 0.1 ms |
| Wall element length | 2 mm |
| Tube wall heat capacity $c_w$ | 830 J.kg$^{-1}$.K$^{-1}$ |
| Tube wall density $\rho_w$ | 2230 kg.m$^{-3}$ |
| Tube wall thermal conductivity $\lambda_w$ | 1.4 W.m$^{-1}$.K$^{-1}$ |
| Temperature of environment $T_{air}$ | 25°C |

A comparison with the simulations performed simultaneously on many parameters shows a good agreement with the experiment (Figure 7). The qualitative trend of the pressure oscillation around a constant average value is reproduced well. The repeating pattern of chaotic duration consists of strong oscillations following a slow decrease that corresponds to the PHP stopover phase.

The simulations show that at stopovers, long liquid plugs covering several evaporator sections (and, accordingly, long bubbles) are formed. The mechanism of the oscillation restart can be described as follows. Sometimes bubbles nucleate inside the liquid plugs thanks to the pressure $p$ decay, during which the wall superheating $T_w - T_{sat}(p)$ increases. When this difference attains $\Delta T_{nucl}$ at some position $x$ inside a plug where $T_w$ is maximal, a bubble nucleates there. If this bubble is close to the extremity of the parent liquid plug, the created "child" liquid plug is small. The bubble expands, which causes a fast child plug motion while the parent plug remains nearly immobile (which justifies the "stopover" name) because of the overall momentum conservation. When the child plug moves over the dry tube wall, its receding meniscus deposits the liquid film, which causes the plug volume reduction and, ultimately, its disappearance. This process has been observed with a high-speed camera in the experiment.

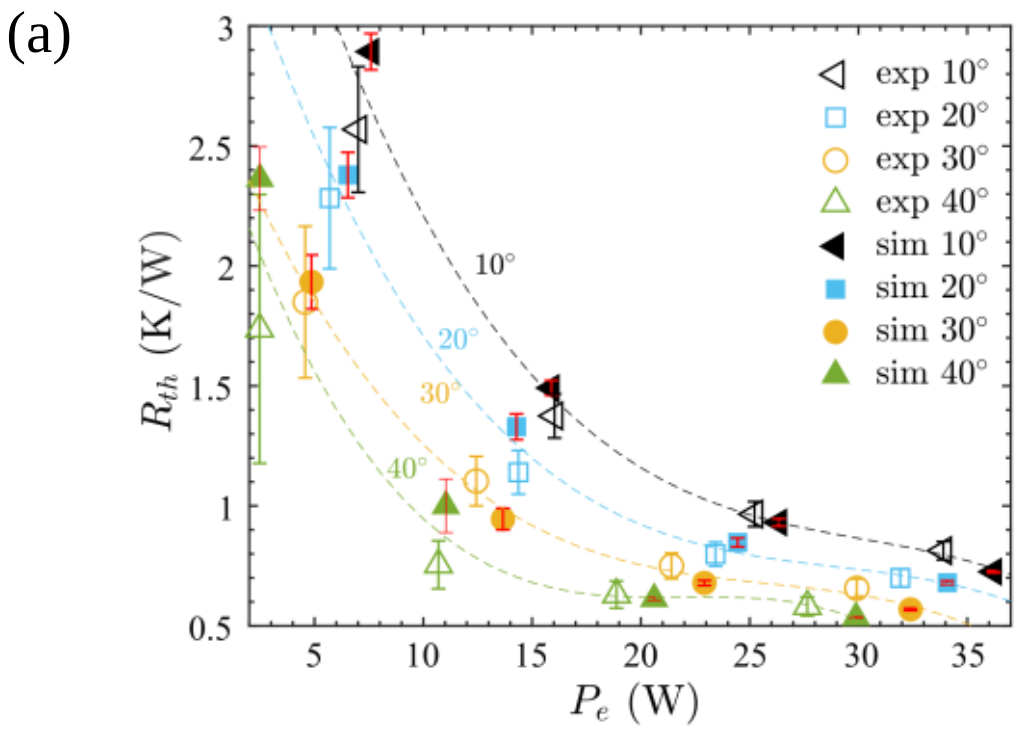

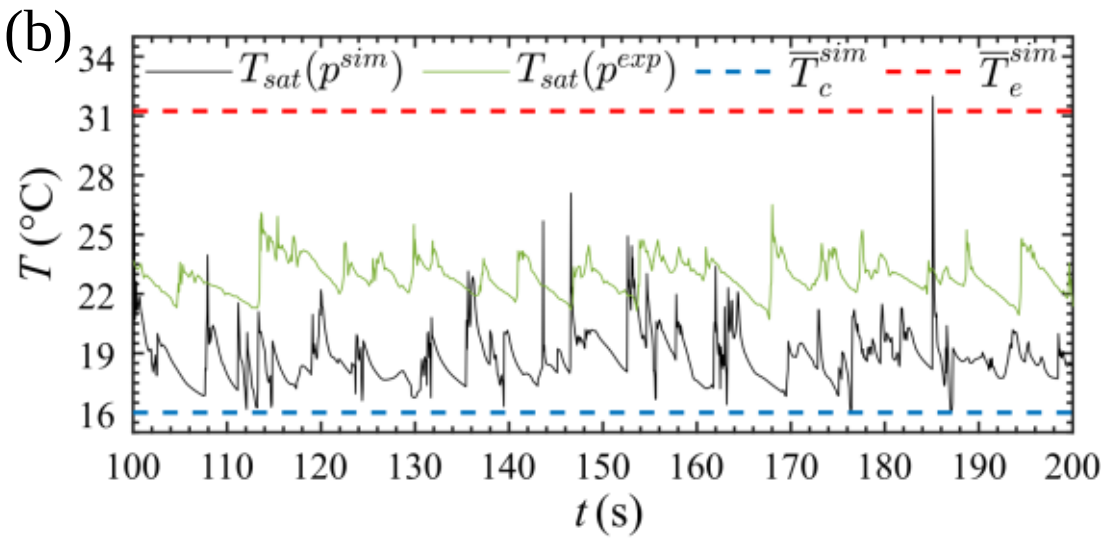

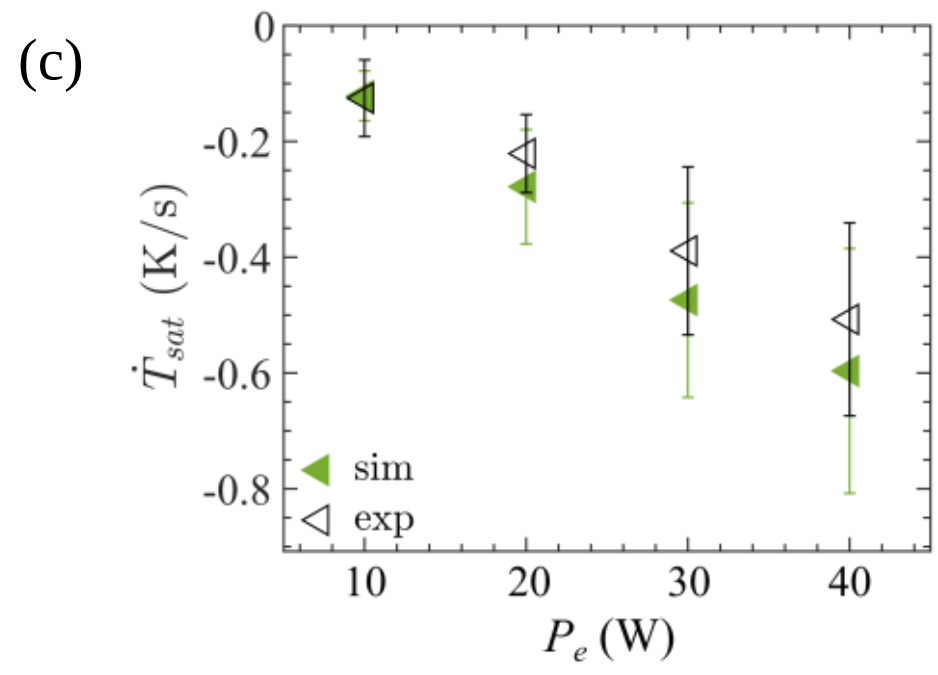

**Figure 7.** Comparison of the experiment (exp) and simulations (sim) for the Smart Loop [19] . (a) Thermal resistance $R_{th}$. The parameter of the curves is the cool plate temperature $T_c$ in °C. (b) Time evolution of the pressure measured in a corner joint A for $T_c$ =10°C and $P_e$= 40 W (depicted in terms of the saturation temperature $T_{sat}$). (c) $T_{sat}$ relaxation rate during stopovers for the same parameters.

When the pressure further decays, a bubble can nucleate deeper inside the plug, which causes a detachment of a child plug with a size comparable to the parent. The expansion of the nucleated bubble can cause in this case the motion of parent plug, which leads to an imbalance of pressures in different bubbles and thus restarting of the global plug motion.

It is evident that the pressure evolution reflects mainly the mass exchange inside the PHP with the pressure increase or decrease corresponding to evaporation or condensation, respectively. In this case, there is a net vapor condensation because its pressure decays. The decay rate $\dot{T}_{sat}$ is defined as an average of $dT_{sat}/dt$ during the whole stopover interval.

Figure 7c shows that $\dot{T}_{sat}$ decreases with $P_e$, which might seem paradoxical. Indeed, at a higher power one can expect a stronger film evaporation, which ought to cause a *slower* pressure decay because the film evaporation counters the overall condensation. This paradox is explained by the film shrinking, i.e., the dry spot growth. At a higher power it is faster, which leads to a weaker vapor generation during stopovers and to a more rapid pressure decay. As discussed above, the pressure decay causes the oscillation restart after a stopover thanks to the bubble nucleation. Therefore, a faster decay leads to a faster oscillation restart, which results in a better performance, i.e., the reduction of $R_{th}$ with $P_e$ in the stopover regime. As the oscillation regime is always stopover at a high $P_e$ according to the regime map in Figure 4c, this should apply to any PHP. Indeed, this feature is commonly observed (see e.g. Figures 2,7a) but was not understood up to now.

## 4. Conclusions

In this paper we discuss several issues. First, it is the experimental validation of the code CASCO v. 4 that is now being introduced for industrial use. It is validated here against two experimental devices. For one of them, only the global parameter validation is presented. For another, we discuss the validation both on the global and local parameters.

The CASCO simulation allows us to study two different functioning regimes, those of continuous oscillations and stopover. The regime map is presented. It is shown that the continuous oscillations occur within a middle range of evaporator powers; the stopover regime occurs both at small and high powers.

The mechanism of the oscillation restart after a stopover is analyzed. An explanation of the reduction of the PHP thermal resistance with the heat load observed generically at high loads is proposed. This explanation is coherent with the PHP flow visualization.

## 5. Acknowledgements

V.S.N. and M.A. acknowledge a financial support of CNES awarded through GdR MFA. M.A. and S.B. thank SPEC/CEA and ST/CNES, respectively, for their hospitality. This work has been partially funded by the grant FA9550-19-S-0003 of European Office of Aerospace Research and Development – Air Force Office of Scientific Research (USA) – ID number WS00317867.